\documentclass[oneside,english]{amsart}
\usepackage[T1]{fontenc}
\usepackage[latin9]{inputenc}
\usepackage{array}
\usepackage{float}
\usepackage{booktabs}
\usepackage{mathtools}
\usepackage{amsthm}
\usepackage{stackrel}

\makeatletter

\providecommand{\tabularnewline}{\\}
\floatstyle{ruled}
\newfloat{algorithm}{tbp}{loa}
\providecommand{\algorithmname}{Algorithm}
\floatname{algorithm}{\protect\algorithmname}

\numberwithin{equation}{section}
\numberwithin{figure}{section}
\theoremstyle{plain}
\newtheorem{thm}{\protect\theoremname}
\theoremstyle{remark}
\newtheorem{rem}[thm]{\protect\remarkname}

\makeatother

\usepackage{babel}
\usepackage{listings}
\providecommand{\remarkname}{Remark}
\providecommand{\theoremname}{Theorem}

\begin{document}
\title{Timing Excess Returns \\
A cross-universe approach to alpha}
\author{Marc Rohloff, Alexander Vogt}
\begin{abstract}
We present a simple model that uses time series momentum in order
to construct strategies that systematically outperform their benchmark.
The simplicity of our model is elegant: We only require a benchmark
time series and several related investable indizes, not requiring
regression or other models to estimate our parameters. \\
\\
We find that our one size fits all approach delivers significant outperformance
in both equity and bond markets while meeting the ex-ante risk requirements,
nearly doubling yearly returns vs. the MSCI World and Bloomberg Barclays
Euro Aggregate Corporate Bond benchmarks in a long-only backtest.
We then combine both approaches into an absolute return strategy by
benchmarking vs. the Eonia Total Return Index and find significant
outperformance at a sharpe ratio of 1.8.\\
\\
Furthermore, we demonstrate that our model delivers a benefit versus
a static portfolio with fixed mean weights, showing that timing of
excess return momentum has a sizeable benefit vs. static allocations.
This also applies to the passively investable equity factors, where
we outperform a static factor exposure portfolio with statistical
significance.\\
\\
Also, we show that our model delivers an alpha after deducting transaction
costs.
\end{abstract}

\thanks{We wish to thank Frank Leyers, without whom this paper would not have
been written}

\maketitle
\pagebreak{}

\tableofcontents{}

\pagebreak{}

\section{Introduction}

\subsection{State of the research}

Time series momentum is a long studied effect finding time stability
of excess returns across a wide range of asset classes\cite{moskowitz2012time}.
The most common application is to equity factors, relying on the timing
of excess returns associated with the latter in order to construct
an optimal portfolio. There is a wide range of models and research
on factor momentum, most show a significant excess return with respect
to the benchmark. Some models rely on time series information only
\cite{dichtl2019optimal,gupta2019factor,leippold2019fama}, while
others include macro data \cite{hodges2017factor,kinlaw2019crowded}.
There has also been research linking excess factor returns to industry
excess returns \cite{arnott2019factor}.

\subsection{What we do differently}

There is a fundamental aspect separating our approach from the others:
Simplicity. We do not rely on anything but the time series of the
indizes across a look-back period $T$, a certain rebalancing frequency
and a target tracking error for our portfolio, denoted by $\sigma$
in the course of this paper. The simplicity has a profound advantage:
Without having to delve into the specifics of the asset class and
without having to use any assumption of return and volatility models,
we can compute an optimal allocation solely based on time series data.
Furthermore, we demonstrate that our risk targets are well met.

\newpage{}

\section{Methodology}

In order to construct our portfolios, we use the following simple
approach: Take several indices $X_{i}(t)$,$1\leq i\leq n$ and a
benchmark $X_{0}(t)$. Consider the historic excess total return of
the indizes:
\[
R_{i}\coloneqq\frac{X_{i}}{X_{0}}
\]
and the associated return:
\[
\alpha_{i}\coloneqq\frac{dR_{i}}{dt}
\]
Next, take an arbitrary time period $T$ and define the average excess
return:
\[
\Delta_{i}^{T}(t)\coloneqq\frac{1}{T}\int_{t-T}^{t}\alpha_{i}(s)ds
\]
and the covariance coefficients:
\[
\Omega_{ij}^{T}(t)\coloneqq\frac{1}{T}\int_{t-T}^{t}\alpha_{i}(s)\alpha_{j}(s)ds-\Delta_{i}^{T}(t)\Delta_{j}^{T}(t)
\]
For simplicity, define:
\begin{eqnarray*}
\Omega_{0i}^{T} & = & \Omega_{i0}^{T}\\
 & = & 0\\
\\
\Delta_{0}^{T} & = & 0
\end{eqnarray*}

\begin{rem}
Consider a portfolio of weights $P^{T}=(x_{0}^{T},...,x_{n}^{T})$
. For given boundaries $M_{i}^{u}\geq0,M_{i}^{l}\leq0,i\geq1$ denote
the set of admissible weights $S_{M}$ by 
\[
S_{M,\sigma}\coloneqq\left\{ (x_{i})|\stackrel[i=0]{n}{\sum}x_{i}=1,M_{i}^{l}\leq x_{i}\leq M_{i}^{u}\forall i\geq1,\underset{i,j}{\sum}x_{i}\Omega_{ij}^{T}x_{j}\leq\sigma^{2}\right\} 
\]
Then there exists a unique portfolio $P^{T}$ satisfying: 
\begin{eqnarray*}
P^{T}\cdot\Delta^{T} & = & \underset{x\in S_{M,\sigma}}{\max}\left\{ x\cdot\Delta^{T}\right\} \\
 & \eqqcolon & m_{T}\\
\\
\left(P^{T}\right)^{t}\cdot\Omega^{T}\cdot P^{T} & = & \underset{x\in S_{M,\sigma}}{\min}\left\{ x|x\cdot\Delta^{T}=m_{T}\right\} 
\end{eqnarray*}
\end{rem}

This portfolio maximises the historic return over past time period
$T$ while having minimal tracking error below $\sigma$ and satisfying
the allocation restrictions $M$.

\newpage{}

\section{Universes}

As our model is quite general and only requires the notion of a benchmark
and several related indizes, we show that it can be applied to two
very different asset classes to generate excess returns in comparison
to the benchmark, irrespective of their different nature: Equity and
bonds. We implement above algorithm in Python using the cvxpy library
\cite{cvxpy} and consider weights $M=\{(M_{i}^{l}=0,M_{i}^{u}=1)\}$
and $T=91$ days for all backtests. 

\subsection{Equity}

The first application of our timing model, equity, has a natural set
of indizes related to the standard market benchmark: Factor indizes.
As factors intrinsically carry an excess return \cite{asness2019quality,carhart1997persistence,dolvin2017has,fama1988dividend,fama1995size,shi2014size,zhang2005value},
one can ask themselves whether timing their allocation provides a
benefit with regards to the benchmark. \\
\\
For these backtests, we consider the MSCI Benchmarks Europe, World
and USA and add to each benchmark five factor indizes: Minimum Volatility,
Momentum, Size, Value and Quality. As we are interested in the effect
of rebalancing on the portfolio, we conduct two backtests for each
benchmark, one with a rebalancing frequency of seven days, and one
with 28 days between rebalancing dates. As we base our backtest on
daily data, the backtest period starts the first year all indizes
have daily data available. We fix the tracking error $\sigma=4\%$,
which is a tracking error typically seen in actively managed portfolios
and define our sets of benchmarks and indizes as follows.

\subsubsection{World}

The World subuniverse consists of the MSCI World Benchmark and the
following indizes (we also list their Bloomberg tickers):
\begin{enumerate}
\item Benchmark: MSCI World / NDDUWI Index
\item Min. Vol.: MSCI World Minimum Volatility Index / M00IWO\$O Index
\item Momentum: MSCI World Momentum Index / M1WOMOM Index
\item Size: MSCI World Size Tilt Index / M1WOMEQ Index
\item Value: MSCI World Enhanced Value Index / M1WOEV Index
\item Quality: MSCI World Sector Neutral Quality Index / M1WONQ Index
\end{enumerate}
The World universe has the same backtest period as the US universe.

\subsubsection{US}

The US subuniverse consists of the MSCI USA Benchmark and the following
indizes:
\begin{enumerate}
\item Benchmark: MSCI USA / NDDUUS Index
\item Min. Vol.: MSCI USA Minimum Volatility Index / M1USMVOL Index
\item Momentum: MSCI USA Momentum Index / M1US000\$ Index
\item Size: MSCI USA Size Tilt Index / M1CXBRG Index
\item Value: MSCI USA Enhanced Value Index / M1USEV Index
\item Quality: MSCI USA Sector Neutral Quality Index / M1USSNQ Index
\end{enumerate}
The backtest in the US universe starts on 2000-01-07 and ends on 2020-01-14
due to the daily data for all factors being available from 1999 onwards.

\subsubsection{Europe}

The Europe subuniverse consists of the MSCI Europe Benchmark and the
following indizes:
\begin{enumerate}
\item Benchmark: MSCI Europe / MSDEE15N Index
\item Min. Vol.: MSCI Europe Minimum Volatility Index / MAEUVOE Index
\item Momentum: MSCI Europe Momentum Index / MAEUMMT Index
\item Size: MSCI Europe Size Tilt Index / M7EUMEW Index
\item Value: MSCI Europe Enhanced Value Index / M7EUEV Index
\item Quality: MSCI Europe Sector Neutral Quality Index / M7ESNQ Index
\end{enumerate}
The Europe universe has a backtest running from 2003-01-07 till 2020-01-14,
as unlike the other two universes the daily data is available from
2002 onwards.

\newpage{}

\subsection{Bonds}

The second universe of assets we look at is bonds. Unlike equity,
there is no readily investable factor universe for bonds, albeit there
having been recent research about bond factors \cite{houweling2017factor,correia2012value}.
For timing to work, though, in theory, we only need indizes that differ
predictevely from the benchmark. In order to provide access to systematic
performance deviation from the benchmark, we select from a pool of
systematic indizes that have a broad range of performance drivers:
inflation, securization, credit risk and interest rate risk. We choose
a rebalancing frequency of 28 days and a tracking error of 2\%. 

\subsubsection{European Bonds}

The Europe universe consists of the Bloomberg Barclays Euro Aggregate
Corporate Bond Index Benchmark and the following indizes:
\begin{enumerate}
\item Benchmark: Bloomberg Barclays Euro Aggregate Corporate Bond Index
/ LECPTREU Index
\item Covered Bonds: Bloomberg Barclays Securitized - Covered Bond Index
/ LSC1TREU Index
\item Government Bonds: Barclays EuroAgg Treasury Index / LEATTREU Index
\item Inflation Linked Bonds: Bloomberg Barclays Euro Govt Inflation-Linked
Bond All Maturities Index / BEIG1T Index
\item Long Duration: Bloomberg Barclays Euro Government 30 Year Term Index
/ BCEX1T Index
\item Short Duration: Bloomberg Barclays Euro-Aggregate Government 1-3 Year
Index / LEG1TREU Index
\item High Yield: Bloomberg Barclays Pan-European High Yield / LP01TREU
Index
\end{enumerate}
As with the equity backtests, our backtests starts Jan. 7th on the
first year having daily data for all indizes, which is 2005-01-07.

\newpage{}

\subsection{Absolute Return}

As a fun excercise demonstrating the simplicity of our model, we combine
above indizes into an absolute return strategy: We again use $T=91$
days, rebalance monthly and set our target tracking error to $\sigma=2\%$
to the Eonia TR benchmark, use a both long only approach as well as
a backtest with $M=\{(M_{0}^{l}=0,M_{0}^{u}=1),(M_{i}^{l}=-1,M_{i}^{u}=1)|i>0\}$,
i.e. a long-short approach, and define our universe as follows:

\subsubsection{Absolute Return}

The Europe universe consists of the Bloomberg Barclays Euro Aggregate
Corporate Bond Index Benchmark and the following indizes:
\begin{enumerate}
\item Benchmark: Eonia Total Return Index / DBDCONIA Index
\item World: MSCI World / NDDUWI Index
\item Min. Vol.: MSCI World Minimum Volatility Index / M00IWO\$O Index
\item Momentum: MSCI World Momentum Index / M1WOMOM Index
\item Size: MSCI World Size Tilt Index / M1WOMEQ Index
\item Value: MSCI World Enhanced Value Index / M1WOEV Index
\item Quality: MSCI World Sector Neutral Quality Index / M1WONQ Index
\item Corp. Bonds: Bloomberg Barclays Euro Aggregate Corporate Bond Index
/ LECPTREU Index
\item Covered Bonds: Bloomberg Barclays Securitized - Covered Bond Index
/ LSC1TREU Index
\item Government Bonds: Barclays EuroAgg Treasury Index / LEATTREU Index
\item Inflation Linked Bonds: Bloomberg Barclays Euro Govt Inflation-Linked
Bond All Maturities Index / BEIG1T Index
\item Long Duration: Bloomberg Barclays Euro Government 30 Year Term Index
/ BCEX1T Index
\item Short Duration: Bloomberg Barclays Euro-Aggregate Government 1-3 Year
Index / LEG1TREU Index
\item High Yield: Bloomberg Barclays Pan-European High Yield / LP01TREU
Index
\end{enumerate}
The start date for the backtest is the same as for the bonds one due
to sharing all indizes, i.e. 2005-01-07.

\newpage{}

\section{Results}

We report our data based on daily arithmetic returns and annualize
the figures. We compute the following statistics:
\begin{itemize}
\item Volatility (VOL) (mean daily volatility $\cdot\sqrt{365.25}$)
\item Return, annualized (i.e mean daily return $\cdot365.25$)
\item Sharpe Ratio (SR), the ratio of return to volatility
\item Alpha (mean daily difference between portfolio and benchmark, annualized)
\item Tracking Error (TE), the standard deviation of alpha
\item Information Ratio (IR): alpha divided by tracking error
\item Maximum relative drawdown (MRDD): The maximum relative drawdown of
the portfolio with respect to the benchmark. If the MRDD is -15\%,
the portfolio has a negative alpha of -15\% from the highest point
to the lowest relative to the benchmark.
\item TER: Our total expense ratio per annum. It is an estimate based on
an index bid/ask spread of 5 bp.
\end{itemize}

\subsection{Equity universes}

\subsubsection{World}

Both strategies examined have a significant information ratio with
respect to the benchmark, having a p-value of $0.000028$. Also, the
observed volatility is similar to the benchmark, albeit with a 68\%
higher annual return. The tracking error metrics are comparable to
those of the factors, whereas the MRDD is strikingly lower. The target
tracking error of 4\% is reasonably attained with 4.5\% realized tracking
error.\\

\begin{center}
\noindent\resizebox{\textwidth}{!}{%
\begin{tabular*}{13.5cm}{@{\extracolsep{\fill}}>{\centering}m{1cm}>{\centering}p{1cm}>{\centering}p{1cm}>{\centering}p{1cm}>{\centering}p{1cm}>{\centering}p{1cm}>{\centering}p{1cm}>{\centering}m{1cm}>{\centering}p{1cm}}
\toprule 
\raggedright{} & \raggedright{}{\scriptsize{}Benchmark} & \begin{raggedright}
{\scriptsize{}28 Day Rebala}{\scriptsize\par}
\par\end{raggedright}
\raggedright{}{\scriptsize{}ncing} & \raggedright{}{\scriptsize{}7 Day Rebalancing} & \raggedright{}{\scriptsize{}Min. Vol.} & \raggedright{}{\scriptsize{}Momentum} & \raggedright{}{\scriptsize{}Quality} & \raggedright{}{\scriptsize{}Size} & \raggedright{}{\scriptsize{}Value}\tabularnewline
\midrule
\midrule 
\raggedright{}{\scriptsize{}Return} & \raggedright{}{\scriptsize{}5.6\%} & \raggedright{}{\scriptsize{}9.5\%} & \raggedright{}{\scriptsize{}9.4\%} & \raggedright{}{\scriptsize{}7.1\%} & \raggedright{}{\scriptsize{}7.7\%} & \raggedright{}{\scriptsize{}7.0\%} & \raggedright{}{\scriptsize{}7.3\%} & \raggedright{}{\scriptsize{}8.2\%}\tabularnewline
\midrule 
\raggedright{}{\scriptsize{}VOL} & \raggedright{}{\scriptsize{}16.3\%} & \raggedright{}{\scriptsize{}15.6\%} & \raggedright{}{\scriptsize{}15.6\%} & \raggedright{}{\scriptsize{}12.8\%} & \raggedright{}{\scriptsize{}17.1\%} & \raggedright{}{\scriptsize{}16.5\%} & \raggedright{}{\scriptsize{}15.2\%} & \raggedright{}{\scriptsize{}16.4\%}\tabularnewline
\midrule 
\raggedright{}{\scriptsize{}SR} & \raggedright{}{\scriptsize{}0.34} & \raggedright{}{\scriptsize{}0.61} & \raggedright{}{\scriptsize{}0.60} & \raggedright{}{\scriptsize{}0.56} & \raggedright{}{\scriptsize{}0.45} & \raggedright{}{\scriptsize{}0.43} & \raggedright{}{\scriptsize{}0.48} & \raggedright{}{\scriptsize{}0.50}\tabularnewline
\midrule 
\raggedright{} & \raggedright{} & \raggedright{} & \raggedright{} & \raggedright{} & \centering{} & \raggedright{} & \raggedright{} & \raggedright{}\tabularnewline
\midrule 
\raggedright{}{\scriptsize{}Alpha} & \raggedright{}{\scriptsize{}---} & \raggedright{}{\scriptsize{}4.0\%} & \raggedright{}{\scriptsize{}4.0\%} & \raggedright{}{\scriptsize{}1.1\%} & \raggedright{}{\scriptsize{}1.6\%} & \raggedright{}{\scriptsize{}1.1\%} & \raggedright{}{\scriptsize{}1.3\%} & \raggedright{}{\scriptsize{}1.9\%}\tabularnewline
\midrule 
\raggedright{}{\scriptsize{}TE} & \raggedright{}{\scriptsize{}---} & \raggedright{}{\scriptsize{}4.5\%} & \raggedright{}{\scriptsize{}4.4\%} & \raggedright{}{\scriptsize{}6.8\%} & \raggedright{}{\scriptsize{}7.5\%} & \raggedright{}{\scriptsize{}3.1\%} & \raggedright{}{\scriptsize{}5.3\%} & \raggedright{}{\scriptsize{}5.7\%}\tabularnewline
\midrule 
\raggedright{}{\scriptsize{}IR} & \raggedright{}{\scriptsize{}---} & \raggedright{}\textbf{\scriptsize{}0.90} & \raggedright{}\textbf{\scriptsize{}0.90} & \raggedright{}{\scriptsize{}0.17} & \raggedright{}{\scriptsize{}0.21} & \raggedright{}{\scriptsize{}0.34} & \raggedright{}{\scriptsize{}0.24} & \raggedright{}{\scriptsize{}0.34}\tabularnewline
\midrule 
\raggedright{} & \raggedright{} & \raggedright{} & \raggedright{} & \raggedright{} & \raggedright{} & \raggedright{} & \raggedright{} & \raggedright{}\tabularnewline
\midrule 
\raggedright{}{\scriptsize{}MRDD} & \raggedright{}{\scriptsize{}---} & \raggedright{}{\scriptsize{}-8.5\%} & \raggedright{}{\scriptsize{}-7,1\%} & \raggedright{}{\scriptsize{}-17.6\%} & \raggedright{}{\scriptsize{}-20.7\%} & \raggedright{}{\scriptsize{}-10.9\%} & \raggedright{}{\scriptsize{}-14.6\%} & \raggedright{}{\scriptsize{}-25.3\%}\tabularnewline
\bottomrule
\end{tabular*}}\\
~\\
~\\
\par\end{center}

Examining the mean allocations per index, one can see that based on
the mean allocation, one would expect a mean excess return of around
1.4\% based on average allocation statistics, implying the 2.6\% additional
excess return is an active contribution from timing. We also tested
the performance of the portfolio vs. an equal weighted portfolio,
not finding any significant difference between the performance vs.
mean or vs. equal weighted portfolios, hence we omit the latter. Further
examined is the strategy alpha vs. a static strategy possessing the
same mean allocation (``Mean''). The outperformance of the timing
strategy vs. the mean allocation is significant with a p-value of
$0.011304$.\\

\begin{center}
{\footnotesize{}}%
\begin{tabular}{cccc}
\toprule 
 & {\footnotesize{}Index} & {\footnotesize{}28 Day} & {\footnotesize{}7 Day}\tabularnewline
\midrule
{\footnotesize{}Mean Weights} & {\footnotesize{}Benchmark} & {\footnotesize{}6.5\%} & {\footnotesize{}6.4\%}\tabularnewline
\cmidrule{2-4} \cmidrule{3-4} \cmidrule{4-4} 
 & {\footnotesize{}Min. Vol.} & {\footnotesize{}16.0\%} & {\footnotesize{}15.3\%}\tabularnewline
\cmidrule{2-4} \cmidrule{3-4} \cmidrule{4-4} 
 & {\footnotesize{}Momentum} & {\footnotesize{}25.9\%} & {\footnotesize{}25.4\%}\tabularnewline
\cmidrule{2-4} \cmidrule{3-4} \cmidrule{4-4} 
 & {\footnotesize{}Quality} & {\footnotesize{}15.6\%} & {\footnotesize{}16.0\%}\tabularnewline
\cmidrule{2-4} \cmidrule{3-4} \cmidrule{4-4} 
 & {\footnotesize{}Size} & {\footnotesize{}10.0\%} & {\footnotesize{}10.6\%}\tabularnewline
\cmidrule{2-4} \cmidrule{3-4} \cmidrule{4-4} 
 & {\footnotesize{}Value} & {\footnotesize{}26.0\%} & {\footnotesize{}26.1\%}\tabularnewline
\midrule 
 &  &  & \tabularnewline
\midrule 
{\footnotesize{}Allocation} &  & {\footnotesize{}1.4\%} & {\footnotesize{}1.4\%}\tabularnewline
\midrule 
{\footnotesize{}Active} &  & {\footnotesize{}2.6\%} & {\footnotesize{}2.6\%}\tabularnewline
\midrule 
 &  &  & \tabularnewline
\midrule 
{\footnotesize{}TER} &  & {\footnotesize{}0.57\%} & {\footnotesize{}1.15\%}\tabularnewline
\midrule 
{\footnotesize{}Turnover} &  & {\footnotesize{}1147\%} & {\footnotesize{}2299\%}\tabularnewline
\midrule 
 &  &  & \tabularnewline
\midrule 
{\footnotesize{}Alpha vs. Mean} &  & {\footnotesize{}1.9\%} & {\footnotesize{}1.8\%}\tabularnewline
\midrule 
{\footnotesize{}TE vs. Mean} &  & {\footnotesize{}3.7\%} & {\footnotesize{}3.8\%}\tabularnewline
\midrule 
{\footnotesize{}IR vs. Mean} &  & \textbf{\footnotesize{}0.51} & \textbf{\footnotesize{}0.53}\tabularnewline
\bottomrule
\end{tabular}{\footnotesize{}}\\
{\footnotesize{}~}\\
{\footnotesize{}~}\\
\par\end{center}

With a TER of 0.57\%, our 28-day strategy is still well viable after
trading costs whereas the 7-day strategy would perform signifanctly
worse.

\newpage{}

\subsubsection{US}

Same as in the world universe, the outperformance of the strategy
with respect to the benchmark is significant with a p-value of $0.004145$.
Again, the volatility is similar to that of the benchmark with a 38\%
higher annual return, the further picture is similar as well: A tracking
error that well matches the 4\% target and a lower MRDD than the factors
themselves.\\

\noindent \begin{center}
\noindent\resizebox{\textwidth}{!}{%
\begin{tabular*}{13.5cm}{@{\extracolsep{\fill}}>{\centering}m{1cm}>{\centering}p{1cm}>{\centering}p{1cm}>{\centering}p{1cm}>{\centering}p{1cm}>{\centering}p{1cm}>{\centering}p{1cm}>{\centering}p{1cm}>{\centering}p{1cm}}
\toprule 
\raggedright{} & \raggedright{}{\scriptsize{}Benchmark} & \begin{raggedright}
{\scriptsize{}28 Day Rebala}{\scriptsize\par}
\par\end{raggedright}
\raggedright{}{\scriptsize{}ncing} & \raggedright{}{\scriptsize{}7 Day Rebalancing} & \raggedright{}{\scriptsize{}Min. Vol.} & \raggedright{}{\scriptsize{}Momentum} & \raggedright{}{\scriptsize{}Quality} & \raggedright{}{\scriptsize{}Size} & \raggedright{}{\scriptsize{}Value}\tabularnewline
\midrule
\midrule 
\raggedright{}{\scriptsize{}Return} & {\scriptsize{}7.0\%} & {\scriptsize{}9.6\%} & {\scriptsize{}10.1\%} & {\scriptsize{}8.1\%} & {\scriptsize{}9.4\%} & {\scriptsize{}7.7\%} & {\scriptsize{}9.8\%} & {\scriptsize{}9.8\%}\tabularnewline
\midrule 
\raggedright{}{\scriptsize{}Vol.} & {\scriptsize{}20.6\%} & {\scriptsize{}20.2\%} & {\scriptsize{}20.0\%} & \noindent \centering{}{\scriptsize{}17.3\%} & {\scriptsize{}21.0\%} & {\scriptsize{}19.8\%} & {\scriptsize{}22.4\%} & {\scriptsize{}21.5\%}\tabularnewline
\midrule 
\raggedright{}{\scriptsize{}SR} & \raggedright{}{\scriptsize{}0.34} & \raggedright{}{\scriptsize{}0.47} & {\scriptsize{}0.51} & {\scriptsize{}0.47} & {\scriptsize{}0.45} & {\scriptsize{}0.39} & {\scriptsize{}0.44} & {\scriptsize{}0.46}\tabularnewline
\midrule 
\raggedright{} & \raggedright{} & \raggedright{} & \raggedright{} & \raggedright{} & \raggedright{} & \raggedright{} & \raggedright{} & \centering{}\tabularnewline
\midrule 
\raggedright{}{\scriptsize{}Alpha} & \raggedright{}{\scriptsize{}---} & {\scriptsize{}2.5\%} & {\scriptsize{}3.1\%} & {\scriptsize{}0.8\%} & {\scriptsize{}1.7\%} & {\scriptsize{}0.5\%} & {\scriptsize{}2.0\%} & {\scriptsize{}2.0\%}\tabularnewline
\midrule 
\raggedright{}{\scriptsize{}TE} & \raggedright{}{\scriptsize{}---} & {\scriptsize{}4.3\%} & {\scriptsize{}4.2\%} & {\scriptsize{}6.4\%} & {\scriptsize{}7.4\%} & {\scriptsize{}3.2\%} & {\scriptsize{}6.4\%} & {\scriptsize{}5.0\%}\tabularnewline
\midrule 
\raggedright{}{\scriptsize{}IR} & \raggedright{}{\scriptsize{}---} & \textbf{\scriptsize{}0.59} & \textbf{\scriptsize{}0.75} & {\scriptsize{}0.13} & {\scriptsize{}0.23} & {\scriptsize{}0.14} & {\scriptsize{}0.31} & {\scriptsize{}0.40}\tabularnewline
\midrule 
\raggedright{} & \raggedright{} &  &  &  &  &  &  & \tabularnewline
\midrule 
\raggedright{}{\scriptsize{}MRDD} & \raggedright{}{\scriptsize{}---} & {\scriptsize{}-9.4\%} & {\scriptsize{}-7.1\%} & {\scriptsize{}-15.2\%} & {\scriptsize{}-23.0\%} & {\scriptsize{}-10.5\%} & {\scriptsize{}-18.9\%} & {\scriptsize{}-18.2\%}\tabularnewline
\bottomrule
\end{tabular*}}\\
~\\
~\\
\par\end{center}

\begin{flushleft}
The other characteristics of the US universe match the pattern found
in world: An allocation contribution to excess return that is lower
than the observes excess return, leaving an active contribution of
roughly equal magnitude. However, the outperformance vs. a static
portfolio of same mean weight is not statistically significant. Interestingly,
after costs, the 28-day and 7-day strategy have approximately the
same alpha.\\
\par\end{flushleft}

\noindent \begin{center}
{\footnotesize{}}%
\begin{tabular}{cccc}
\toprule 
 & {\footnotesize{}Index} & {\footnotesize{}28 Day} & {\footnotesize{}7 Day}\tabularnewline
\midrule
{\footnotesize{}Mean Weights} & {\footnotesize{}Benchmark} & {\footnotesize{}6.8\%} & {\footnotesize{}5.7\%}\tabularnewline
\cmidrule{2-4} \cmidrule{3-4} \cmidrule{4-4} 
 & {\footnotesize{}Min. Vol.} & {\footnotesize{}15.2\%} & {\footnotesize{}15.3\%}\tabularnewline
\cmidrule{2-4} \cmidrule{3-4} \cmidrule{4-4} 
 & {\footnotesize{}Momentum} & {\footnotesize{}22.9\%} & {\footnotesize{}21.1\%}\tabularnewline
\cmidrule{2-4} \cmidrule{3-4} \cmidrule{4-4} 
 & {\footnotesize{}Quality} & {\footnotesize{}13.6\%} & {\footnotesize{}14.2\%}\tabularnewline
\cmidrule{2-4} \cmidrule{3-4} \cmidrule{4-4} 
 & {\footnotesize{}Size} & {\footnotesize{}20.6\%} & {\footnotesize{}20.4\%}\tabularnewline
\cmidrule{2-4} \cmidrule{3-4} \cmidrule{4-4} 
 & {\footnotesize{}Value} & {\footnotesize{}21.0\%} & {\footnotesize{}23.2\%}\tabularnewline
\midrule 
 &  &  & \tabularnewline
\midrule 
{\footnotesize{}Allocation} &  & {\footnotesize{}1.4\%} & {\footnotesize{}1.4\%}\tabularnewline
\midrule 
{\footnotesize{}Active} &  & {\footnotesize{}1.1\%} & {\footnotesize{}1.7\%}\tabularnewline
\midrule 
 &  &  & \tabularnewline
\midrule 
{\footnotesize{}TER} &  & {\footnotesize{}0.59\%} & {\footnotesize{}1.26\%}\tabularnewline
\midrule 
{\footnotesize{}Turnover} &  & {\footnotesize{}1172\%} & {\footnotesize{}2501\%}\tabularnewline
\midrule 
 &  &  & \tabularnewline
\midrule 
{\footnotesize{}Alpha vs. Mean} &  & {\footnotesize{}0.7\%} & {\footnotesize{}1.3\%}\tabularnewline
\midrule 
{\footnotesize{}TE vs. Mean} &  & {\footnotesize{}3.7\%} & {\footnotesize{}3.7\%}\tabularnewline
\midrule 
{\footnotesize{}IR vs. Mean} &  & {\footnotesize{}0.19} & {\footnotesize{}0.35}\tabularnewline
\bottomrule
\end{tabular}{\footnotesize{}\newpage}{\footnotesize\par}
\par\end{center}

\subsubsection{Europe}

The pattern repeats. We have a significant outperformance (p-value
of $0.021692$) with respect to the benchmark, at similar volatility.
The tracking error is comparable to the factors again, with only Quality
having a lower MRDD than the strategy.\\

\begin{center}
\noindent\resizebox{\textwidth}{!}{%
\begin{tabular*}{13.5cm}{@{\extracolsep{\fill}}>{\centering}m{1cm}>{\centering}p{1cm}>{\centering}p{1cm}>{\centering}p{1cm}>{\centering}p{1cm}>{\centering}p{1cm}>{\centering}p{1cm}>{\centering}p{1cm}>{\centering}p{1cm}}
\toprule 
\raggedright{} & \raggedright{}{\scriptsize{}Benchmark} & \begin{raggedright}
{\scriptsize{}28 Day Rebala}{\scriptsize\par}
\par\end{raggedright}
\raggedright{}{\scriptsize{}ncing} & \raggedright{}{\scriptsize{}7 Day Rebalancing} & \raggedright{}{\scriptsize{}Min. Vol.} & \raggedright{}{\scriptsize{}Momentum} & \raggedright{}{\scriptsize{}Quality} & \raggedright{}{\scriptsize{}Size} & \raggedright{}{\scriptsize{}Value}\tabularnewline
\midrule
\midrule 
\raggedright{}{\scriptsize{}Return} & {\scriptsize{}8.2\%} & {\scriptsize{}10.3\%} & {\scriptsize{}10.0\%} & {\scriptsize{}9.0\%} & {\scriptsize{}11.3\%} & {\scriptsize{}9.9\%} & {\scriptsize{}9.8\%} & {\scriptsize{}9.4\%}\tabularnewline
\midrule 
\raggedright{}{\scriptsize{}Vol.} & {\scriptsize{}18.1\%} & {\scriptsize{}17.5\%} & {\scriptsize{}17.2\%} & {\scriptsize{}13.7\%} & {\scriptsize{}17.5\%} & {\scriptsize{}17.4\%} & {\scriptsize{}18.2\%} & {\scriptsize{}19.6\%}\tabularnewline
\midrule 
\raggedright{}{\scriptsize{}SR} & {\scriptsize{}0.45} & {\scriptsize{}0.59} & {\scriptsize{}0.58} & {\scriptsize{}0.66} & {\scriptsize{}0.64} & {\scriptsize{}0.57} & {\scriptsize{}0.54} & {\scriptsize{}0.48}\tabularnewline
\midrule 
\raggedright{} & \raggedright{} & \raggedright{} & \raggedright{} & \raggedright{} & \raggedright{} & \raggedright{} & \raggedright{} & \raggedright{}\tabularnewline
\midrule 
\raggedright{}{\scriptsize{}Alpha} & \raggedright{}{\scriptsize{}---} & {\scriptsize{}2.4\%} & {\scriptsize{}2.1\%} & {\scriptsize{}0.6\%} & {\scriptsize{}2.2\%} & {\scriptsize{}1.2\%} & {\scriptsize{}1.2\%} & {\scriptsize{}0.8\%}\tabularnewline
\midrule 
\raggedright{}{\scriptsize{}TE} & \raggedright{}{\scriptsize{}---} & {\scriptsize{}4.3\%} & {\scriptsize{}4.2\%} & {\scriptsize{}6.1\%} & {\scriptsize{}7.3\%} & {\scriptsize{}3.5\%} & {\scriptsize{}5.5\%} & {\scriptsize{}4.2\%}\tabularnewline
\midrule 
\raggedright{}{\scriptsize{}IR} & \raggedright{}{\scriptsize{}---} & \textbf{\scriptsize{}0.55} & \textbf{\scriptsize{}0.49} & {\scriptsize{}0.1} & {\scriptsize{}0.31} & {\scriptsize{}0.34} & {\scriptsize{}0.21} & {\scriptsize{}0.2}\tabularnewline
\midrule 
\raggedright{} & \raggedright{} & \raggedright{} & \raggedright{} & \raggedright{} & \raggedright{} & \raggedright{} & \raggedright{} & \raggedright{}\tabularnewline
\midrule 
\raggedright{}{\scriptsize{}MRDD} & \raggedright{}{\scriptsize{}---} & {\scriptsize{}-9.4\%} & {\scriptsize{}-8.3\%} & {\scriptsize{}-16.5\%} & {\scriptsize{}-21.1\%} & {\scriptsize{}-8.0\%} & {\scriptsize{}-20.2\%} & {\scriptsize{}-22.4\%}\tabularnewline
\bottomrule
\end{tabular*}}\\
~\\
~\\
\par\end{center}

For the European universe, there still is an active component to excess
return. However, it is not statistically significant. We find that
the alpha ex costs of the 7-day strategy is nearly half that of the
28-day strategy.\\

\begin{center}
{\footnotesize{}}%
\begin{tabular}{cccc}
\toprule 
 & {\footnotesize{}Index} & {\footnotesize{}28 Day} & {\footnotesize{}7 Day}\tabularnewline
\midrule
{\footnotesize{}Mean Weights} & {\footnotesize{}Benchmark} & {\footnotesize{}4.3\%} & {\footnotesize{}4.3\%}\tabularnewline
\cmidrule{2-4} \cmidrule{3-4} \cmidrule{4-4} 
 & {\footnotesize{}Min. Vol.} & {\footnotesize{}16.7\%} & {\footnotesize{}17.4\%}\tabularnewline
\cmidrule{2-4} \cmidrule{3-4} \cmidrule{4-4} 
 & {\footnotesize{}Momentum} & {\footnotesize{}24.3\%} & {\footnotesize{}24.1\%}\tabularnewline
\cmidrule{2-4} \cmidrule{3-4} \cmidrule{4-4} 
 & {\footnotesize{}Quality} & {\footnotesize{}15.7\%} & {\footnotesize{}15.0\%}\tabularnewline
\cmidrule{2-4} \cmidrule{3-4} \cmidrule{4-4} 
 & {\footnotesize{}Size} & {\footnotesize{}17.1\%} & {\footnotesize{}16.3\%}\tabularnewline
\cmidrule{2-4} \cmidrule{3-4} \cmidrule{4-4} 
 & {\footnotesize{}Value} & {\footnotesize{}21.8\%} & {\footnotesize{}22.9\%}\tabularnewline
\midrule 
 &  &  & \tabularnewline
\midrule 
{\footnotesize{}Allocation} &  & {\footnotesize{}1.2\%} & {\footnotesize{}1.2\%}\tabularnewline
\midrule 
{\footnotesize{}Active} &  & {\footnotesize{}1.2\%} & {\footnotesize{}0.9\%}\tabularnewline
\midrule 
 &  &  & \tabularnewline
\midrule 
{\footnotesize{}TER} &  & {\footnotesize{}0.53\%} & {\footnotesize{}1.19\%}\tabularnewline
\midrule 
{\footnotesize{}Turnover} &  & {\footnotesize{}1257\%} & {\footnotesize{}2781\%}\tabularnewline
\midrule 
 &  &  & \tabularnewline
\midrule 
{\footnotesize{}Alpha vs. Mean} &  & {\footnotesize{}0.5\%} & {\footnotesize{}0.2\%}\tabularnewline
\midrule 
{\footnotesize{}TE vs. Mean} &  & {\footnotesize{}3.7\%} & {\footnotesize{}3.6\%}\tabularnewline
\midrule 
{\footnotesize{}IR vs. Mean} &  & {\footnotesize{}0.14} & {\footnotesize{}0.06}\tabularnewline
\bottomrule
\end{tabular}{\footnotesize{}\newpage}{\footnotesize\par}
\par\end{center}

\subsection{Bonds}

The result for the bond universe is stunning. The strategy has an
alpha of equal magnitude to that of the benchmark (p-value $<0.00001$)
and a small maximal relative drawdown of 5.3\%, the target tracking
error of 2.0\% is almost reached.\\

\noindent \begin{center}
\noindent\resizebox{\textwidth}{!}{%
\begin{tabular*}{13.5cm}{@{\extracolsep{\fill}}>{\centering}m{1cm}>{\centering}p{1cm}>{\centering}p{1cm}>{\centering}p{1cm}>{\centering}p{1cm}>{\centering}p{1cm}>{\centering}p{1cm}>{\centering}p{1cm}>{\centering}p{1cm}}
\toprule 
\raggedright{} & \raggedright{}{\scriptsize{}Benchmark} & \begin{raggedright}
{\scriptsize{}28 Day Rebala}{\scriptsize\par}
\par\end{raggedright}
\raggedright{}{\scriptsize{}ncing} & \raggedright{}{\scriptsize{}Covered Bonds } & \raggedright{}{\scriptsize{}Government Bonds } & \raggedright{}{\scriptsize{}Inflation Linked Bonds } & \raggedright{}{\scriptsize{}Long Duration } & \raggedright{}{\scriptsize{}Short Duration } & {\scriptsize{}High Yield}\tabularnewline
\midrule
\midrule 
\raggedright{}{\scriptsize{}Return} & {\scriptsize{}3.7\%} & {\scriptsize{}7.1\%} & {\scriptsize{}3.7\%} & {\scriptsize{}4.1\%} & {\scriptsize{}2.7\%} & {\scriptsize{}6.9\%} & {\scriptsize{}2.0\%} & {\scriptsize{}7.5\%}\tabularnewline
\midrule 
\raggedright{}{\scriptsize{}Vol.} & {\scriptsize{}2.5\%} & {\scriptsize{}3.4\%} & {\scriptsize{}2.1\%} & {\scriptsize{}3.8\%} & {\scriptsize{}5.0\%} & {\scriptsize{}8.5\%} & {\scriptsize{}1.1\%} & {\scriptsize{}5.0\%}\tabularnewline
\midrule 
\raggedright{}{\scriptsize{}SR} & {\scriptsize{}1.47} & {\scriptsize{}2.09} & {\scriptsize{}1.74} & {\scriptsize{}1.09} & {\scriptsize{}0.55} & {\scriptsize{}0.80} & {\scriptsize{}1.83} & {\scriptsize{}1.50}\tabularnewline
\midrule 
\raggedright{} & \raggedright{} &  &  &  &  &  &  & \tabularnewline
\midrule 
\raggedright{}{\scriptsize{}Alpha} & \raggedright{}{\scriptsize{}---} & {\scriptsize{}3.4\%} & {\scriptsize{}-1.0\%} & {\scriptsize{}0.3\%} & {\scriptsize{}-0.7\%} & {\scriptsize{}2.3\%} & {\scriptsize{}-1.2\%} & {\scriptsize{}2.8\%}\tabularnewline
\midrule 
\raggedright{}{\scriptsize{}TE} & \raggedright{}{\scriptsize{}---} & {\scriptsize{}2.4\%} & {\scriptsize{}1.3\%} & {\scriptsize{}2.7\%} & {\scriptsize{}4.1\%} & {\scriptsize{}7.1\%} & {\scriptsize{}2.2\%} & {\scriptsize{}4.9\%}\tabularnewline
\midrule 
\raggedright{}{\scriptsize{}IR} & \raggedright{}{\scriptsize{}---} & \textbf{\scriptsize{}1.37} & {\scriptsize{}-0.76} & {\scriptsize{}0.11} & {\scriptsize{}-0.17} & {\scriptsize{}0.32} & {\scriptsize{}-0.57} & {\scriptsize{}0.56\%}\tabularnewline
\midrule 
\raggedright{} & \raggedright{} &  &  &  &  &  &  & \tabularnewline
\midrule 
\raggedright{}{\scriptsize{}MRDD} & \raggedright{}{\scriptsize{}---} & {\scriptsize{}-5.3\%} & {\scriptsize{}-16.4\%} & {\scriptsize{}-17.6\%} & {\scriptsize{}-26.3\%} & {\scriptsize{}-16.2\%} & {\scriptsize{}-35.3\%} & {\scriptsize{}-37.8\%}\tabularnewline
\bottomrule
\end{tabular*}}\\
~\\
~\\
\par\end{center}

If one looks at the mean allocation, there is a 40\% High Yield quota.
However, the risk profile of the strategy seems nowhere near that
of High Yield, its volatility and tracking error are markedly lower.
The MRDD is drastically smaller - at only 5.3\% compared to the 37.8\%
drawdown of the High Yield index compared to the benchmark during
the Global Financial Crisis. The active component of the return is
sizeable and with a cost of 0.48\% p.a. the strategy would still yield
an alpha of 2.9\%. We have a significant outperformance vs. a static
portfolio allocation at a p-value of $0.0005$.\\

\begin{center}
{\footnotesize{}}%
\begin{tabular}{ccc}
\toprule 
 & {\footnotesize{}Index} & {\footnotesize{}28 Day}\tabularnewline
\midrule
{\footnotesize{}Mean Weights} & {\footnotesize{}Benchmark} & {\footnotesize{}13.4\%}\tabularnewline
\cmidrule{2-3} \cmidrule{3-3} 
 & {\footnotesize{}Covered Bonds} & {\footnotesize{}10.5\%}\tabularnewline
\cmidrule{2-3} \cmidrule{3-3} 
 & {\footnotesize{}Government Bonds} & {\footnotesize{}6.1\%}\tabularnewline
\cmidrule{2-3} \cmidrule{3-3} 
 & {\footnotesize{}Inflation Linked Bonds} & {\footnotesize{}8.0\%}\tabularnewline
\cmidrule{2-3} \cmidrule{3-3} 
 & {\footnotesize{}Long Duration} & {\footnotesize{}14.3\%}\tabularnewline
\cmidrule{2-3} \cmidrule{3-3} 
 & {\footnotesize{}Short Duration} & {\footnotesize{}7.7\%}\tabularnewline
\cmidrule{2-3} \cmidrule{3-3} 
 & {\footnotesize{}High Yield} & {\footnotesize{}40.0\%}\tabularnewline
\midrule 
 &  & \tabularnewline
\midrule 
{\footnotesize{}TER} &  & {\footnotesize{}0.48\%}\tabularnewline
\midrule 
{\footnotesize{}Turnover} &  & {\footnotesize{}965\%}\tabularnewline
\midrule 
 &  & \tabularnewline
\midrule 
{\footnotesize{}Allocation} &  & {\footnotesize{}1.2\%}\tabularnewline
\midrule 
{\footnotesize{}Active} &  & {\footnotesize{}2.2\%}\tabularnewline
\midrule 
 &  & \tabularnewline
\midrule 
{\footnotesize{}Alpha vs. Mean} &  & {\footnotesize{}1.6\%}\tabularnewline
\midrule 
{\footnotesize{}TE vs. Mean} &  & {\footnotesize{}2.4\%}\tabularnewline
\midrule 
{\footnotesize{}IR vs. Mean} &  & \textbf{\footnotesize{}0.51}\tabularnewline
\bottomrule
\end{tabular}{\footnotesize{}\newpage}{\footnotesize\par}
\par\end{center}

\subsection{Absolute Return}

The strategy works once more. Both long short and long only have a
significant information ratio (p-value $<0.00001$). The volatility
of the long only strategy is much closer to the target volatility
of $2\%$, the difference might be due to a large absolute position
exposure of the long short strategy.\\

\begin{center}
\begin{tabular*}{6.5cm}{@{\extracolsep{\fill}}>{\centering}p{1.5cm}>{\centering}p{1cm}>{\centering}p{1cm}>{\centering}p{1cm}}
\toprule 
\raggedright{} & \raggedright{}{\small{}Benchmark} & \raggedright{}{\small{}Long Only} & \raggedright{}{\small{}Long Short}\tabularnewline
\midrule
\midrule 
\raggedright{}{\small{}Return} & \raggedright{}{\small{}0.9\%} & \raggedright{}{\small{}5.6\%} & \raggedright{}{\small{}6.9\%}\tabularnewline
\midrule 
\raggedright{}{\small{}VOL} & \raggedright{}{\small{}0.1\%} & \raggedright{}{\small{}2.5\%} & \raggedright{}{\small{}3.2\%}\tabularnewline
\midrule 
\raggedright{}{\small{}SR} & \raggedright{}{\small{}7.70} & \raggedright{}{\small{}2.21} & \raggedright{}{\small{}2.13}\tabularnewline
\midrule 
\raggedright{} & \raggedright{} & \raggedright{} & \raggedright{}\tabularnewline
\midrule 
\raggedright{}{\small{}Alpha} & \raggedright{}{\small{}---} & \raggedright{}{\small{}4.7\%} & \raggedright{}{\small{}6.0\%}\tabularnewline
\midrule 
\raggedright{}{\small{}TE} & \raggedright{}{\small{}---} & \raggedright{}{\small{}2.5\%} & \raggedright{}{\small{}3.2\%}\tabularnewline
\midrule 
\raggedright{}{\small{}IR} & \raggedright{}{\small{}---} & \raggedright{}\textbf{\small{}1.85} & \raggedright{}\textbf{\small{}1.85}\tabularnewline
\midrule 
\raggedright{} & \raggedright{} & \raggedright{} & \raggedright{}\tabularnewline
\midrule 
\raggedright{}{\small{}MRDD} & \raggedright{}{\small{}---} & \raggedright{}{\small{}-8.7\%} & \raggedright{}{\small{}-7.2\%}\tabularnewline
\bottomrule
\end{tabular*}\\
~\\
~\\
\par\end{center}

The mean weights are interesting with a mean negative equity exposure
for the long short strategy and a significant high yield allocation
in both strategies. Despite such a significant high yield exposure,
our maximal drawdown, which peaks during the global financial crisis,
is well under control for an absolute return strategy. However, the
high TER and poorer tracking error render the long short strategy
inattractive in comparison to the long only approach. The outperformance
vs. a static mean weight allocation is significant in the long short
approach (p-value $<0.00001$), whereas it is not signficant for the
long only one. 
\begin{center}
{\footnotesize{}}%
\begin{tabular}{cccc}
\toprule 
 & {\small{}Index} & {\small{}Long Only} & {\small{}Long Short}\tabularnewline
\midrule
{\small{}Mean Weights} & {\small{}Benchmark} & {\small{}17.7\%} & {\small{}60.6\%}\tabularnewline
\cmidrule{2-4} \cmidrule{3-4} \cmidrule{4-4} 
 & {\small{}Corp. Bonds} & {\small{}8.2\%} & {\small{}2.2\%}\tabularnewline
\cmidrule{2-4} \cmidrule{3-4} \cmidrule{4-4} 
 & {\small{}Covered Bonds} & {\small{}10.9\%} & {\small{}14.1\%}\tabularnewline
\cmidrule{2-4} \cmidrule{3-4} \cmidrule{4-4} 
 & {\small{}Government Bonds} & {\small{}3.6\%} & {\small{}9.3\%}\tabularnewline
\cmidrule{2-4} \cmidrule{3-4} \cmidrule{4-4} 
 & {\small{}Inflation Linked Bonds} & {\small{}3.4\%} & {\small{}-2.2\%}\tabularnewline
\cmidrule{2-4} \cmidrule{3-4} \cmidrule{4-4} 
 & {\small{}Long Duration} & {\small{}3.9\%} & {\small{}-3.0\%}\tabularnewline
\cmidrule{2-4} \cmidrule{3-4} \cmidrule{4-4} 
 & {\small{}Short Duration} & {\small{}11.0\%} & {\small{}-14.9\%}\tabularnewline
\cmidrule{2-4} \cmidrule{3-4} \cmidrule{4-4} 
 & {\small{}High Yield} & {\small{}32.0\%} & {\small{}32.2\%}\tabularnewline
\cmidrule{2-4} \cmidrule{3-4} \cmidrule{4-4} 
 & {\small{}World} & {\small{}0.1\%} & {\small{}-21.1\%}\tabularnewline
\cmidrule{2-4} \cmidrule{3-4} \cmidrule{4-4} 
 & {\small{}Min. Vol.} & {\small{}3.7\%} & {\small{}1.6\%}\tabularnewline
\cmidrule{2-4} \cmidrule{3-4} \cmidrule{4-4} 
 & {\small{}Momentum} & {\small{}2.1\%} & {\small{}9.0\%}\tabularnewline
\cmidrule{2-4} \cmidrule{3-4} \cmidrule{4-4} 
 & {\small{}Quality} & {\small{}0.4\%} & {\small{}8.5\%}\tabularnewline
\cmidrule{2-4} \cmidrule{3-4} \cmidrule{4-4} 
 & {\small{}Size } & {\small{}1.0\%} & {\small{}-0.5\%}\tabularnewline
\cmidrule{2-4} \cmidrule{3-4} \cmidrule{4-4} 
 & {\small{}Value} & {\small{}2.2\%} & {\small{}4.4\%}\tabularnewline
\midrule 
 &  &  & \tabularnewline
\midrule 
{\small{}TER} &  & {\small{}0.52\%} & {\small{}2.72\%}\tabularnewline
\midrule 
{\small{}Turnover} &  & {\small{}1047\%} & {\small{}5338\%}\tabularnewline
\midrule 
 &  &  & \tabularnewline
\midrule 
{\small{}Allocation} &  & {\small{}1.7\%} & {\small{}3.9\%}\tabularnewline
\midrule 
{\small{}Active} &  & {\small{}3.0\%} & {\small{}2.1\%}\tabularnewline
\midrule 
 &  &  & \tabularnewline
\midrule 
{\footnotesize{}Alpha vs. Mean} &  & {\footnotesize{}0.8\%} & {\footnotesize{}6.0\%}\tabularnewline
\midrule 
{\footnotesize{}TE vs. Mean} &  & {\footnotesize{}2.2\%} & {\footnotesize{}3.2\%}\tabularnewline
\midrule 
{\footnotesize{}IR vs. Mean} &  & {\footnotesize{}0.36} & \textbf{\footnotesize{}1.88}\tabularnewline
\bottomrule
\end{tabular}{\footnotesize\par}
\par\end{center}

\newpage{}

\section{Conclusion}

To summarize, this paper contributes to the debate whether timing
in Portfolio Management decisions are possible or not. In contrast
to most of the existing studies we use a simple model that uses time
series momentum in order to construct strategies that systematically
outperform their benchmark. The one size fits all model works in both
the equity and bond markets where it was possible to achieve statistically
significant alpha in some cases, and an alpha in all. If you combine
stocks and bonds and measure them against a cash benchmark (Absolute
Return), the results are even better. The approach presented in this
paper has not been discussed in this form until now. It turns out
that price momentum alone can lead to significant results.\\
\\
In contrast to other researchers, we rely only on time series of the
indices of equities and bonds across a look-back period, a certain
rebalancing frequency and a target tracking error for our portfolio.
We were able to demonstrate that the risk targets were met, and, at
the same time, significant results could be achieved compared to the
selected benchmarks. This fact is due in particular to the rotation
model presented which natively includes correlation effects. Although
this approach is burdened by high transaction costs, there are still
results that can largely be classified as significant. This applies
in particular to the strategies with global stocks, bonds and absolute
return.\\
\\
As to our stock results, one might expect that the idiosyncratic risk
of smaller universes contributes to noise and hence lower returns
due to more volatility. The signal for the benchmark MSCI World was
able to achieve better results than would have been the case in isolation
for the USA and Europe, which would be in agreement with that hypothesis.\\
\\
An impressive result was achieved in the area of bonds, which can
be classified as statistically highly significant. This is because
of a high High Yield quota. However, the risk profile of the strategy
seems nowhere near that of High Yield, its volatility and tracking
error are markedly lower. The same is true for Absolute Return. High
Yield Bonds are a core investment in the long only as well as in the
long short strategy.\\
\\
Lastly, across all asset universes, we find significant benefits of
timing vs. static allocations. In fact, in no backtest, we found negative
excess returns of our timing strategy vs. the static allocations.
As our model is easily implementable and can be realized using only
passively investable products such as ETFs, this enables a new set
of ``semi-passive'' allocations that actively time their exposure
with a general model to generate excess returns at a fixed target
volatility. 

\newpage{}

\section{Appendix}

As a sample implementation, we provide our code in the appendix. The
first snippet is an optimiser using cvxpy \cite{cvxpy} to compute
the optimal exposure relative to the benchmark.

\begin{algorithm}[H]
\caption{Excess Return Optimization}

\begin{lstlisting}[language=Python,basicstyle={\scriptsize},breaklines=true]
import numpy,scipy
import cvxpy as cp
from pandas import Series import datetime import factor_optimiser import numpy import math 
#Sample optimiser using cvxpy. We need our excess return vector and the covariance matrix of the excess returns as input, as well as the tracking error and any bounds.
def compute_optimal_portfolio(excess_return_vector,covariance_matrix,tracking_error,lower_bounds=None,upper_bounds=None,set_upper_bounds_to_one=True,max_leverage=1,allow_short=False,no_benchmark=False):
	# We set dim to our number of indizes
	dim=len(excess_return_vector)

	# Define a function to compute the historic return
	def historic_return(x):
		return -1*numpy.dot(excess_return_vector,x).item()
	
	#And use cvxpy to configure our bounds depending on the arguments
	x=cp.Variable(dim)
	constraints=[]
	for i in range(0,dim):
		lower=0
		upper=0
		if (not lower_bounds is None) and (len(lower_bounds)==dim):
			lower=lower_bounds[i]
		if (not lower_bounds is None) and (len(upper_bounds)==dim):
			upper=upper_bounds[i]
		elif set_upper_bounds_to_one:
			upper=1
		if(upper!=lower):
			e=[float(j==i) for j in range(0,dim)]
			constraints.append(e@x>=lower)
			constraints.append(e@x<=upper)
	es=[1 for j in range(0,dim)]

	#If we do not want benchmark exposure, we set the sum to one.
	if not no_benchmark:
		constraints.append(es@x<=1)
		constraints.append(cp.quad_form(x,covariance_matrix)<=tracking_error**2)
	else:
		constraints.append(es@x==1)
	#First optimization calculates the maximally possible return
	opt=cp.Problem(cp.Minimize(-1*(excess_return_vector@x)),constraints)
	opt.solve()
	max_return=-1*opt.value

	#The second optimization finds the portfolio with minimum volatility, if there are several with the same maximum return
	constraints.append(x@excess_return_vector==max_return)
	opt2=cp.Problem(cp.Minimize(cp.quad_form(x,covariance_matrix)),constraints)
	opt2.solve()
	return x.value
\end{lstlisting}
\end{algorithm}

The next snippet then takes that optimal allocation and adds the benchmark
exposure.
\begin{algorithm}[H]
\caption{Optimization With Benchmark}
\begin{lstlisting}[language=Python,basicstyle={\scriptsize},breaklines=true]
# We use pandas to provide time series data as a Series
from pandas import Series
import datetime
import factor_optimiser
import numpy
import math

def compute_return_data(benchmark_time_series,factor_time_series,date,window_length_in_days):
	#Generic function to compute excess return time series and excess return covariance and returns an index for each name
	return alphas,covariance,name_map

# We call the compute_optimal_portfolio function from above code and substract the sum of exposures from 1 to get an allocation that includes the benchmark
def compute_optimal_portfolio_with_benchmark(benchmark_time_series,factor_time_series,date,window_length_in_days,tracking_error,no_benchmark=False):
	#Get the data
	alphas,covariance,name_map=compute_return_data(benchmark_time_series,factor_time_series,date,window_length_in_days)
	#Optimise
	optimal_portfolio=factor_optimiser.compute_optimal_portfolio(alphas,covariance,tracking_error,no_benchmark=no_benchmark)
	allocation={}
	sum=0
	for j in factor_time_series:
		allocation[j]=optimal_portfolio[name_map[j]]
		sum+=optimal_portfolio[name_map[j]]
	#And compute benchmark exposure
	allocation["Benchmark"]=1-sum
	return allocation
\end{lstlisting}
\end{algorithm}

\newpage{}

\bibliographystyle{plain}
\bibliography{Sources/sources}

\newpage{}

\end{document}